\documentclass[superscriptaddress,nofootinbib,twocolumn
]{revtex4-1}

\usepackage{hyperref}
\usepackage{bbm}
\usepackage{amsfonts}
\usepackage{mathrsfs}

\usepackage{amsmath,amssymb}

\usepackage{epsfig}
\usepackage{graphicx}              
\usepackage{url}
\usepackage{hyperref}
\usepackage{float}
\usepackage{pstricks}
\usepackage{color}
\usepackage{multirow}
\usepackage{lipsum}

\newcommand{\be}{\begin{equation}}
\newcommand{\ee}{\end{equation}}
\newcommand{\bea}{\begin{eqnarray}}
\newcommand{\eea}{\end{eqnarray}}
\newcommand{\beq}{\begin{eqnarray}}
\newcommand{\eeq}{\end{eqnarray}}

\newcommand{\msol}{M_\odot} 
\newcommand{\pasa}{PASA}
\newcommand{\mnras}{Mon. Not. Roy. Astr. Soc.}
\newcommand{\aj}{Astr. Journ.}
\newcommand{\ks}[1]{#1}

\newcommand{\twiddle}{{\raise.17ex\hbox{$\scriptstyle\sim$}}}

\begin{document}
\title{Pulsar timing can constrain primordial black holes in the LIGO mass window}
\author{Katelin Schutz}
\email{kschutz@berkeley.edu}
\affiliation{Theoretical Physics Group, Lawrence Berkeley National Laboratory, Berkeley, CA 94720, USA}
\affiliation{Berkeley Center for Theoretical Physics, University of California, Berkeley, CA 94720, USA}
\author{Adrian Liu}
\email{Hubble Fellow}
\affiliation{Department of Astronomy, University of California, Berkeley, CA 94720, USA}
\affiliation{Radio Astronomy Laboratory, University of California, Berkeley, CA 94720, USA}

\begin{abstract}
\noindent  The recent discovery of gravitational waves from merging black holes has generated interest in primordial black holes as a possible component of the dark matter. In this paper, we show that pulsar timing may soon have sufficient data to constrain $1$-$1000\,\msol$ primordial black holes via the non-detection of a third-order Shapiro time delay as the black holes move around the Galactic halo. We present the results of a Monte Carlo simulation which suggests that future data from known pulsars may be capable of constraining the PBH density more stringently than other existing methods in the mass range \twiddle1-30$\,\msol$. We find that timing new pulsars discovered using the proposed Square Kilometre Array may constrain primordial black holes in this mass range to comprise less than \twiddle1-10\% of the dark matter. 
\end{abstract}

\maketitle

%%%%%%%%%%%%%
\section{Introduction}
Cosmology is at an observational crossroads. On the one hand, the specific microphysical description of dark matter (DM) has remained elusive for several decades, with some of the best ideas so far looking increasingly constrained (see Refs. \cite{Strigari20131,Baer20151} and references cited therein). Meanwhile, we have recently seen the stunning confirmation of Einstein's theory of general relativity in the form of the Laser Interferometer Gravitational-Wave Observatory (LIGO) detection of the gravitational wave (GW) event GW150914, opening up a new window into understanding our universe \cite{Abbott:2016blz}. 
Recently, there has been interest in the interpretation of GW150914 as a possible signal of DM. In particular, it has been proposed that LIGO may have discovered DM in the form of two coalescing \twiddle$30\, \msol$ primordial black holes (PBHs) \cite{Clesse:2015wea,Bird:2016dcv,Clesse:2016vqa,Kashlinsky:2016sdv}, \ks{though the specific DM interpretation of this scenario appears to depend on the PBH binary formation mechanism \cite{Sasaki:2016jop}}.

The existence of PBHs was originally proposed in 1974 by Carr and Hawking \cite{Carr:1974nx}, initiating an extensive observational program to constrain their abundance. PBHs are of particular theoretical interest because they have all the relevant properties of dark matter  
and can be produced in the early universe by mechanisms such as hybrid inflation \cite{GarciaBellido:1996qt,Lyth:2010zq}, axion-curvaton inflation \cite{Kawasaki:2012wr}, and cosmic strings \cite{Polnarev:1988dh}. Thus, the presence of PBHs can provide a solution to the longstanding mystery of DM and can also shed light on physics at energies well above terrestrially accessible scales. 

The abundance of PBHs with masses $1$-$1000~\msol$ (encompassing the LIGO mass sensitivity window) has already been constrained. Below \twiddle$20 \, \msol$, the fraction of DM comprised of PBHs is constrained by microlensing \cite{Allsman:2000kg, 2011MNRAS.416.2949W,Tisserand:2006zx}. At masses above \twiddle$ 100 \, \msol$, a significant presence of PBHs would disrupt wide stellar binaries \cite{2014ApJ...790..159M, Yoo:2003fr,2009MNRAS.396L..11Q}. At intermediate masses, accretion onto PBHs would induce CMB spectral distortions and modify the spectrum of CMB anisotropies \cite{Ricotti:2007au, Chen:2016pud}, though some have recently argued that 
constraints based on this approach may require significant reinterpretation due to uncertainties in the modeling of complex astrophysical processes \cite{Bird:2016dcv,Clesse:2016vqa}. 
More recently, the survival of stellar clusters in Eradinus~II and other ultra-faint dwarf galaxies has constrained the PBH abundance for masses above \twiddle$10\,\msol$ \cite{Brandt:2016aco}, reinforcing the exclusion of pure, single-mass PBH DM in the entire LIGO mass window. PBHs from inflation are expected to have an extended mass spectrum owing to critical collapse, and the breadth of the mass spectrum could soften the constraints on the total PBH DM abundance \cite{Carr:2016drx, Clesse:2016vqa}. However the typical log-normal mass spectrum from critical collapse is still excluded when the constraints from the observables mentioned above are jointly analyzed \cite{Green:2016xgy}. Strong gravitational lensing of fast radio bursts has been proposed as a means to further constrain the PBH abundance in this mass range with upcoming surveys \cite{Munoz:2016tmg}. 

In this paper, we propose that pulsar timing be used to improve limits on PBH DM in the LIGO mass window. Pulsars are magnetized neutron stars that emit regular pulses of light; in particular, millisecond pulsars (MSPs) have periods whose extreme stability rivals that of atomic clocks \cite{2012MNRAS.427.2780H}. The precise nature of the pulse arrival times makes MSPs ideal for probing general relativistic effects. For example, a search for low-frequency gravitational waves is currently being carried out by monitoring a large collection of MSPs as part of three pulsar timing arrays (PTAs) \cite{Manchesteretal2013,2009arXiv0909.1058J, 2008AIPC..983..633J}. Already, PTAs have been used to place constraints on the existence of nearby supermassive black hole binaries via the non-detection of their GW signatures \cite{Schutz:2015pza, Arzoumanian:2014gja, Babak:2015lua, Zhu:2014rta}. 

Pulsar timing has previously been suggested as a means to detect GWs from PBHs in a scenario where the super-Hubble scalar fluctuations that source the PBHs also generate a second-order tensor perturbation \cite{Saito:2008jc,Bugaev:2010bb}. One can also search for Doppler shifts in the pulse arrival times from accelerations of the pulsar or Earth due to nearby PBHs imparting a gravitational kick \cite{1538-4357-659-1-L33,2012MNRAS.426.1369K}.
Here, we propose using MSPs to measure the Shapiro time delay from PBHs moving relative to the pulsar line of sight (LOS). This motion causes the gravitational potential to change along the LOS and induces a long-term shift in the pulse arrival times. Shapiro delays have been previously explored as a method for detecting DM substructure \cite{Baghram:2011is, Siegel:2007fz, Siegel:2008dw, Clark:2015sha}. The time delay from other compact objects such as stars and astrophysically-generated black holes has also been considered \cite{1996PASA...13..236W, Desai:2015hua, Larchenkova:2007ax,Larchenkova:2009hr}.  Amid any possible skepticism regarding the modeling of complex astrophysical phenomena to derive constraints on PBH DM, pulsar timing poses a relatively clean, independent method for constraint.

The rest of the paper is organized as follows. Following the methods originally outlined in Ref.~\cite{Siegel:2007fz}, we describe the observational setup and signatures of PBH DM in Section~\ref{sec:setup}. We perform an exploratory Monte Carlo analysis to forecast the constraints that could be achieved with pulsar timing from individual PBH motion in Section~\ref{sec:cons}. Concluding remarks follow in Section~\ref{sec:conclusions}.

\section{Observational Signatures of PBHs}
\label{sec:setup}
 \begin{figure}[t!]
\includegraphics[width=0.48\textwidth]{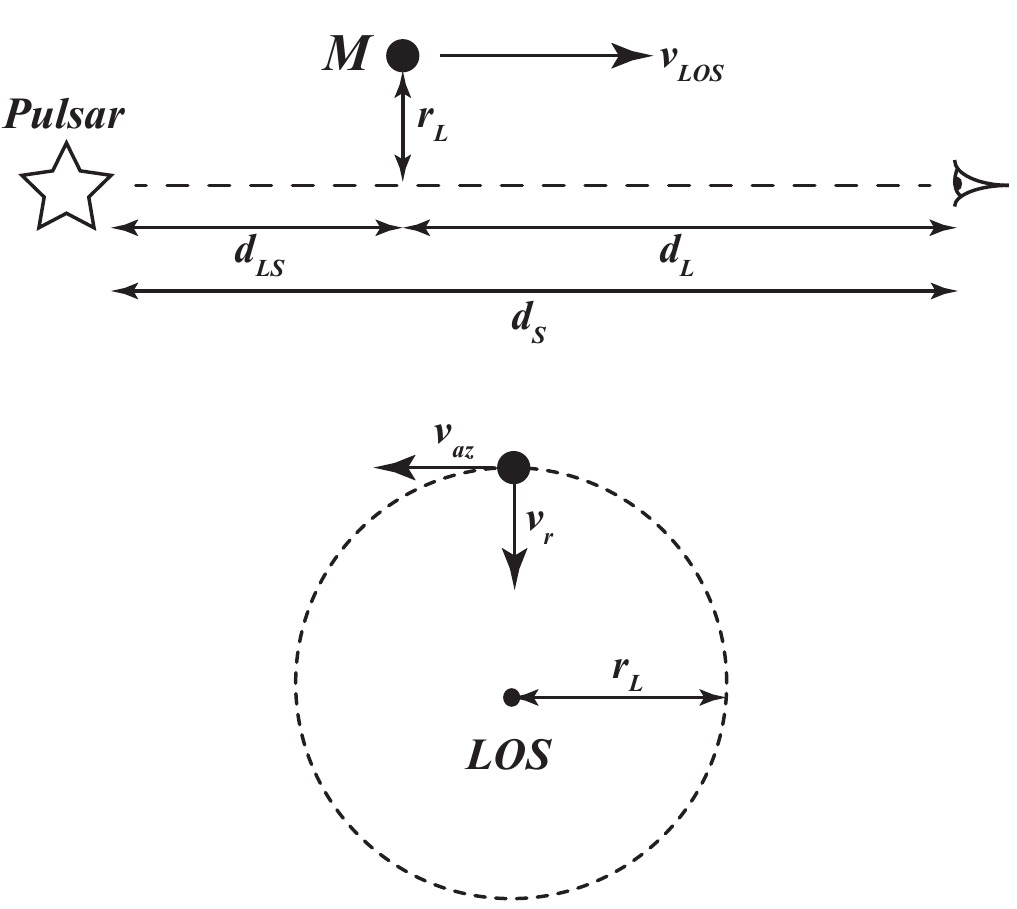}
\caption{Schematic observational setup for a single pulsar. The top diagram shows the geometry from a side view, while the bottom shows a cross section from the point of view of the observer.}
\label{fig:diagram}
\end{figure}

Suppose the pulse times of arrival (TOA) from a pulsar are monitored. Standard effects --- such as Doppler shifting from the Earth's orbit around the sun and the pulsar's proper motion --- can influence the TOA. PTA analysis pipelines have proven rather adept at modeling and adjusting for a wide variety of effects in their searches for gravitational waves, and the same techniques can potentially be applied to searches for other general relativistic phenomena. 

In computing their impact on the TOA, we treat PBHs as point masses since a simple order-of-magnitude estimate shows that the odds of observing strong-field influences on the pulse TOA are one in a trillion over a continuous 30-year observing period.\footnote{Suppose we conservatively define strong-field effects to be important if a particular pulsar LOS is within $1000\,r_s$ of a PBH, where $r_s \equiv 2GM/c^2$ is its Schwartzschild radius. For a pulsar located $d_S\,$\twiddle$\,1\,\textrm{kpc}$ away from Earth, its LOS will thus have a cross-section for strong-field events  given by \twiddle$1000\,r_s d_S$. This cross-section is exposed to a PBH wind of $n\,vT$, where $n\,$\twiddle$\,10^{-2} (\msol / M)\,\textrm{pc}^{-3}$ is the number density of PBHs (assuming a mean density of $10^{-2} \, \msol$/pc$^3$ in the local Galactic halo), $v\,$\twiddle$\,220\,\textrm{km/s}$ is the characteristic velocity of PBHs, and $T$ is the observing time. The total number of expected strong-field events is thus \twiddle$ 1000\,r_s  n\,vT d_S$ which is \twiddle$ 10^{-12}$ for $T=30\,\textrm{years}$.} 
In the weak-field limit, a point mass affects the spacetime metric as \beq ds^2 = - \left(1 + 2 \frac{\phi}{c^2}\right) c^2 dt^2 + \left(1- 2 \frac{\phi}{c^2}\right)  d\mathbf{x}^{2} \eeq
where $c$ is the speed of light, $\phi\equiv -GM / r$ is the Newtonian potential of a point with mass $M$ located a distance $r$ away, and $G$ is the gravitational constant. At leading order in $\phi/c^2$, two terms source delayed TOA for a null geodesic in this gauge. The first is known as the Shapiro time delay and comes from the time-time part of the metric \cite{shapiro1964fourth}. The second is a geometric time delay, coming from spatial curvature in the path of the null geodesic. Since we are concerned with PBHs in the mass range $1$-$1000~\msol$ that have thus far evaded detection from methods that involve geometric shifts (\emph{i.e.} microlensing) we ignore subdominant geometric contributions to the time delay. Thus, the relevant quantity is the cumulative Shapiro delay along the undeflected null geodesic,
\begin{align}\label{time_delay}  \Delta t &= -\frac{2}{c^3} \int \phi \, ds = \frac{2 G M}{c^3} \int_{-d_L}^{d_{LS}} \frac{1}{\sqrt{r_L^2 + z^2}}\, dz \\& =  \frac{2 G M}{c^3} \ln\left(\frac{\sqrt{r_L^2 + d_{LS}^2} +d_{LS}}{\sqrt{r_L^2 + d_L^2} - d_L}\right) \nonumber\end{align}
where the integration variable extends along the LOS and where the geometry of the system is defined in Figure~\ref{fig:diagram}. Specifically, $r_L$ is the radial distance from the LOS to the PBH lens, $d_S$ is the distance from the Earth to the pulsar, $d_L$ is the distance along the LOS from the Earth to the PBH, and $d_{LS}$ is the distance along the LOS between the source and the lens.

If the point mass is stationary relative to the LOS, its presence can never be measured using the Shapiro delay, as this would amount to a constant phase shift in the pulse TOA. Thus, pulsar timing is not sensitive to the total foreground mass, but only to \emph{changes} in time delay as the foreground mass moves, analogous to the integrated Sachs-Wolfe effect in cosmology \cite{Sachs:1967er}. We therefore promote the geometric parameters in Eq.~\eqref{time_delay} to be functions of the observing time $T$, and the Shapiro delay can then be measured in principle from analyzing the TOA. 

In practice, observational degeneracies will place even more restrictions on how much of a measured time delay can be attributed to a Shapiro delay. To see this, consider an example where a PBH is initially located at $d_L = d_{LS} \equiv d_0$ along the LOS of a pulsar. Further assume that the PBH begins at $r_L = r_{L,0}$ and moves at a constant inward radial velocity, $v_r$. Under this simple example where $v_{LOS}$ and $v_{az}$ (the initial LOS and azimuthal velocities from Fig.~\ref{fig:diagram}) are zero, the arrival time $t(n)$ of the $n$th pulse from the pulsar is given by
\begin{align}
\label{eq:expansion}
t(n) &\approx n P + \frac{n^2}{2}  P \dot{P}  \\
&\quad + \frac{2 G M}{c^3} \ln\left(\frac{\sqrt{(r_{L,0}-v_r n P)^2 + d_0^2} +d_0}{\sqrt{(r_{L,0}-v_r n P)^2 + d_0^2} - d_0}\right)\nonumber \\
&\approx  \frac{4 G M}{c^3} \ln\left(\frac{2d_0}{r_{L,0}}\right) + nP \left( 1 + \frac{4 G M  v_r}{c^3 r_{L,0}} \right) \qquad \nonumber \\
&\quad+ \frac{n^2P^2}{2} \left( \frac{\dot{P}}{2P} +\frac{4 G M v_r^2}{c^3 r_{L,0}^2} \right) + n^3 P^3 \frac{4 G Mv_r^3}{3 c^3 r_{L,0}^3} + \dots, \nonumber \qquad \quad
\end{align}
where $P$ and $\dot{P}$ are the initial period of the pulsar and its time derivative, respectively. \ks{In the first equality above, we drop the highly subdominant coupling term between the period evolution and the Shapiro delay.} In the second equality, we expanded in powers of $n$ and additionally assumed that $r_{L,0} \ll d_0$ to simplify this example. Note that we are using a quasi-static approximation where the PBH motion is not factored into the integral because effects from changes to the geometry between pulses on millisecond timescales are negligible.

As discussed above, the $\mathcal{O}(n^0)$ constant term from Eq.~\eqref{eq:expansion} is not observable. However, it is not the only unobservable term: the $\mathcal{O}(n)$ term due to the PBH can be absorbed into one's measurement of $P$, and the $\mathcal{O}(n^2)$ term is degenerate with a redefined $\dot{P}$. Neither $P$ nor $\dot{P}$ can be ascertained for a given pulsar outside of its timing model because we do not have any way to independently measure the intrinsic pulse period or spin-down rate of the pulsar. Therefore, any observable Shapiro delay signature from PBHs must arise from $\mathcal{O}(n^3)$ and higher terms,\footnote{Our omission of the linear term is responsible for the differences between our analysis and analogous proposals for probing DM substructure in Ref. \cite{Clark:2015sha}, which would suggest stronger constraints if applied to the problem of PBHs.} similar to the timing statistic outlined in Ref. \cite{matsakis1997statistic}. Over a long time $T$, the observable portion of the Shapiro delay is given by
\begin{gather}
\Delta t_\textrm{Shapiro} \equiv t ( n = T / P)\hspace{-0.05cm} - t ( n=0) \hspace{-0.05cm} \approx \hspace{-0.05cm} \frac{4 GM v_r^3 T^3}{3 c^3 r_{L,0}^3} \hspace{-0.05cm}+ \ldots \quad\quad \raisetag{1.2\baselineskip}
\label{eq:observable}
\end{gather}
to leading order.
Note that the $T^3$ dependence seen above amounts to an integrated constraint on the second time derivative of the period.
Under the standard assumption of pulsar spin-down due to magnetic dipole braking \cite{1977puls.book.....M}, this second time derivative of the period is expected to be very small and positive, resulting in nanosecond-level delays over a 30-year observing period. Meanwhile, the sign of Eq.~\eqref{eq:observable} can be positive or negative and (as we will show) can lead to timing residuals at an observable level of \twiddle0.1 $\mu$s.
Of course, in real observations there are other ways to generate this time dependence without PBHs, for instance large fluctuations in the magnetic field around the pulsar \cite{Pons:2012cm} or a binary companion with a wide orbit \cite{Kaplan:2016ymq}. Additionally, Shapiro delays can in principle be caused by luminous compact objects along the LOS (like stars). However, these other signatures can ostensibly be searched for in followup observations.

In the next section we simulate just how many pairs of PBHs and pulsars are likely to yield an observable signature, under the assumptions of a particular observing time, pulse timing precision, PBH number density, and velocity distribution. This in turn can be converted into a forecast for how well pulsar timing can probe the abundance of PBH DM.

\section{Projected Constraints}
\label{sec:cons}
In this section, we describe our Monte Carlo simulation of the constraints that could feasibly be achieved using future pulsar timing data. First, we consider a collection of pulsars consisting of all 178 MSPs in the ATNF pulsar catalog with period derivatives between $10^{-21}$ and $10^{-18}$ in magnitude \cite{2005AJ....129.1993M}. %\footnote{For reference, the record stability for an atomic clock is $1.6 \times 10^{-18}$ \cite{hinkley2013atomic}.} 
Most of these MSPs are within 2~kpc of Earth and many have been actively monitored for decades, continuing to be monitored into the future as part of at least one of the three active PTAs. Some of the pulsars in this collection have a very similar LOS, leaving 143 non-redundant pulsar LOS. For a more futuristic projection, we also consider a collection of 2000 pulsars, which could reasonably be discovered and monitored using the Square Kilometre Array (SKA) \cite{Keane:2014vja}. We assume that the SKA pulsars are isotropically distributed at a distance of 2~kpc from Earth.
For both observational setups, we assume a fiducial 30-year observing period. Present-day pulsar timing can achieve optimal  timing precision (the root-mean-square timing residual) of around 0.05 $\mu$s for the best-timed pulsars (see \emph{e.g.} pulsar J0645+5158 in the NANOGrav nine-year data set \cite{2015ApJ...813...65T}), so we optimistically assume a detection threshold of a 0.05~$\mu$s shift over the observing period. 

For each of the two ensembles of pulsars, the simulation populates the local Milky Way (MW) DM halo with PBHs in the region between the pulsars and Earth. For the local DM distribution, we assume a singular isothermal sphere density profile $\rho\sim 1/r^2$, where $r$ is the radius from the Galactic center, consistent with the observed flat rotation curve in the local MW \cite{binney2011galactic}. \ks{We have checked that our results are insensitive to using this profile rather than using a Navarro-Frenk-White profile \cite{Navarro:1996gj} owing to the fact that we are simulating a relatively small volume of the  galactic halo near the Earth at $r\approx8\,\textrm{kpc}$.} To normalize the density profile, we take the local DM density to be $10^{-2} \,\msol$/pc$^3 \sim 0.3$ GeV/cm$^3$ \ks{in order to be roughly consistent with a broad range of measurements \cite{Read:2014qva}}. For the velocities, we assume an isotropic Maxwell-Boltzmann distribution, consistent with this density profile \cite{binney2011galactic}, with an RMS velocity of 220~km/s.

\begin{figure}[t!]
\hspace{-0.4cm}\includegraphics[width=0.5\textwidth]{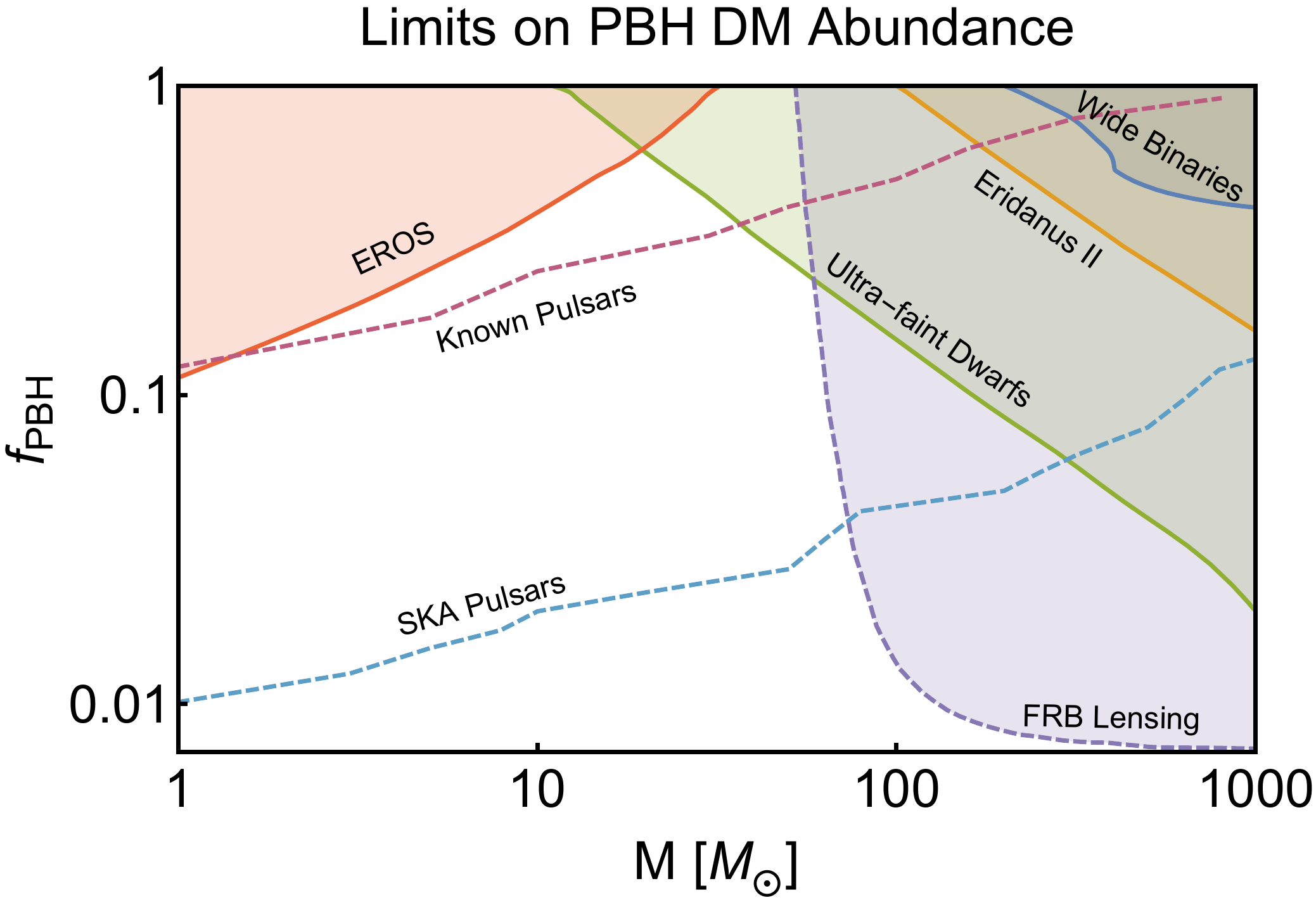}
\caption{Projected constraints from searching for individual motions of PBHs near the 143 distinct LOSs of stable MSPs in the ATNF catalogue (known pulsars) as well as for 2000 MSPs that can be detected and monitored using the SKA \cite{Keane:2014vja}. Also shown are the most stringent observational constraints from microlensing \cite{Tisserand:2006zx}, constraints from ultra-faint dwarf galaxies and Eridanus II assuming conservative dynamical parameters \cite{Brandt:2016aco}, conservative projected constraints from lensing of $10^4$ fast radio bursts \cite{Munoz:2016tmg}, and existing constraints from the non-disruption of wide stellar binaries \cite{2009MNRAS.396L..11Q}. For emphasis, dashed lines denote projected constraints.}\vspace{-0.3cm}
\label{fig:constraints}
\end{figure}

In populating the MW halo with PBHs, we make the standard assumption that the PBHs all have the same mass. For a given PBH mass, the quantity we constrain is the fraction of DM that is made of PBHs, by mass. This quantity, $f_\text{PBH}$, can be expressed as $f_\text{PBH}= \rho_\text{PBH} / \rho_\text{DM}$, 
where $\rho_X$ is the mass density of species $X$.
Each time we populate the local MW halo with PBHs and assign them a velocity, we compute the expected change in the time delay for each pulsar over a 30-year observing period, subtracting off the linear and quadratic pieces due to the aforementioned degeneracies. Note that in the subtraction procedure we use the full expression for the time delay rather than assuming the simplified geometry of Eq.~\eqref{eq:expansion}. Scanning through parameter space, we
simulate MW halos for a given $f_\text{PBH}$ and PBH mass $M$ and record the fraction which contain a PBH that is detectable from having a third-order change in the time delay greater than 0.05 $\mu$s. In Figure~\ref{fig:constraints}, we show the $f_\text{PBH}$ values where 90\% of simulated halos have a detectable PBH of mass $M$.

The trend and normalization of the projected constraint can be explained with a simple order-of-magnitude argument. The time delay is maximized when the LOS radius is small and the PBH velocity is radial because such a configuration maximally changes the potential along the LOS. In this limit, the third-order time delay scales as  $\Delta t_\text{obs}\,$\twiddle$\,G M (v_r T)^3 / ( r_L c)^3$, as shown in Eq.~\eqref{eq:expansion}. This means that for a given observation time, timing precision, fiducial velocity, and PBH mass we can solve for the $r_L$ that would give an observable time delay and define a PBH cylinder of influence with volume $\pi d_S r_L^2$. Then assuming a roughly constant mean DM density of $10^{-2} \,\msol$/pc$^3$ in the local MW halo and assuming most detected events are near the $\Delta t_\text{RMS}$ timing resolution threshold, the expected number $\langle N_\text{PBH} \rangle$ of detectable PBHs within a given pulsar's timing signal is \beq \langle N_\text{PBH} \rangle \sim \pi  d_S \left(\frac{ 2 G M  }{c^3 \Delta t_\text{RMS}}\right)^{2/3}  \frac{v_r^2 T^2 \,f_\text{PBH} \,\rho_\text{DM} }{ M}. \label{eq:oom} \eeq  
Expecting roughly one detectable PBH in any random configuration of the galactic halo with $N_{LOS}$ pulsar LOS yields  
\begin{align} f_\text{PBH} \sim\,& 0.1 \times \left(\frac{M}{\msol}\right)^{1/3} \left(\frac{220 \,\text{km/s}}{v_r}\right)^2 \left(\frac{30 \,\text{years}}{T}\right)^2\nonumber \\
&\times \left(\frac{\Delta t_\text{RMS}}{0.05 \,\mu\text{s}}\right)^{2/3} \left(\frac{2\, \text{kpc}}{\langle d_S \rangle}\right) \left( \frac{100}{N_\text{LOS}}\right). \end{align} 
The rough normalization of $f_\text{PBH}$ and the scaling $f_\text{PBH} \sim M^{1/3} $ can be seen in the simulation results in Figure \ref{fig:constraints}. The plot does not extend past 1$\,\msol$ because at that mass, the radius of influence to the LOS is comparable to the distance traversed over the observing time; in this case, the expansion of Eq.~\eqref{eq:expansion} cannot be truncated, meaning that the $M^{1/3}$ scaling does not persist and sensitivity drops precipitously.

Our projections should be interpreted with the following practical considerations in mind. First, it is interesting to note that since the SKA is projected to provide strong constraints, the full 30-year observing time may not be absolutely necessary and  a shorter observing campaign could be enough to meaningfully improve the constraints. Another consideration regarding these projections arises from the fact that \emph{e.g.} Ref.~\cite{Desai:2015hua} has estimated that the Shapiro delay from stars is a factor of \twiddle20 weaker than for DM, meaning that the level of constraint in these forecasts could be influenced by stellar ``backgrounds'' starting below $f\,$\twiddle$\,0.05$. In the event of a detection of a third-order Shapiro delay, followup observations should be made; from the scaling arguments presented above Eq.~\eqref{eq:oom}, the characteristic angular scale of influence is \twiddle1 arcsecond, making these followups fairly targeted. In the event that stellar backgrounds start to become a limiting factor, the spatial distribution of stars in the Galactic disk means that pulsars at high Galactic latitude could provide a cleaner constraint on PBHs. A final remark is that we have assumed an isotropic distribution of SKA MSPs, despite the fact that many known pulsars lie along a similar LOS. Despite the fact that this reduces the effective volume of space where PBHs can induce Shapiro delays, the redundancy in LOS may actually help boost the signal-to-noise ratio. If a third-order Shapiro delay is coherently seen in multiple pulsars along a particular LOS in a very correlated way, that could dispel any concern that the signal is coming from something locally affecting any individual pulsar, such as a wide binary companion or fluctuations in the magnetic field.

\ks{In addition to the physical considerations discussed above, there are also issues of noise and degeneracy to consider. One possible issue is that there may be some level of degeneracy between a third-order Shapiro delay and other expected effects such as the nonzero first derivative of the pulsar period. This degeneracy arises because $T^2$ and $T^3$ are not orthogonal functions. We have performed a simple Fisher matrix analysis and find that the error in $\Delta t_\textrm{Shapiro}$ (with the degeneracy taken into account) is at the sub-nanosecond level. Another issue comes from red noise signatures which are known to be present in some pulsars \cite{2010ApJ...725.1607S, 2015ApJ...813...65T}. Due to red noise effects, it is possible that some pulsars will not achieve 50~ns timing residuals over the fiducial observing period of 30 years. Given the relatively weak $f_\text{PBH} \sim \Delta t_\text{RMS}^{2/3}$ scaling and known levels of red noise in pulsar timing analysis \cite{2015ApJ...813...65T}, we do not expect our forecasts to change by more than order unity factors. We further emphasize that the Shapiro delay has additional distinguishing features. For example, prewhitened timing residuals in pulsars with red noise fluctuate about zero and can depend on the observed radio frequency, for instance due to unmodelled effects in the interstellar medium \cite{2015ApJ...813...65T}. Meanwhile, in the sensitivity window for the Shapiro delay, a key prediction is that the time delay will be monotonic (\emph{i.e.} the signal strength will grow with time) and achromatic. Therefore, we believe that the Shapiro delay can be observed or excluded with sufficient signal-to-noise even in the presence of red noise due to these distinct observational features of the signal. A detailed followup analysis would further elucidate the possible statistical significance of a detection or constraint.}

\section{Conclusions} 
In this paper, we have shown that pulsar timing has the potential to constrain the abundance of PBH DM. As PBHs move through the Galactic halo, close encounters with the line of sight between the Earth and a given millisecond pulsar will result in third-order Shapiro delays in pulse arrival times. Over observational timescales of 30 years, the cumulative delay can be observed with timing resolutions achieved by modern PTAs using known pulsars as well as undiscovered pulsars, for instance from the proposed SKA.

We find the resulting upper limits on the fraction of PBH DM, $f_\textrm{DM}$, to scale as $M^{1/3}$, where $M$ is the assumed mass of the PBHs. This scaling arises because for a given fixed local DM density, increasing $M$ has the effect of decreasing the number of PBHs in the Galactic halo, reducing the number of close encounters with pulsar LOS. However, a larger $M$ also results in larger Shapiro delays, relaxing the criterion of what constitutes a ``close" encounter. The former effect dominates the latter one, and thus a non-detection of a third-order time delay places the strongest upper limits on $f_\textrm{DM}$ at low values of $M$. 

Figure \ref{fig:constraints} confirms our scaling arguments using numerical simulations with known pulsars and SKA pulsars with optimistic levels of timing resolution. The simulations suggest that PBH DM can be constrained at the 1-10\% level in the mass range $M$\twiddle$1$-$1000 \msol$. Thus, we expect that using future data to perform the analysis proposed in this paper will yield competitive constraints on the abundance of PBHs, providing yet another crucial step in our ongoing quest to understand the nature of DM.

\label{sec:conclusions}
\section*{Acknowledgements}
It is an immense pleasure to thank Dorota Grabowska, Alan Guth, Chung-Pei Ma, Chiara Mingarelli, Christopher Mogni, Siddharth Mishra-Sharma, Scott Ransom, Justin Ripley, Tracy Slatyer, and Leo Stein for useful conversations and correspondence pertaining to this work. \ks{We also wish to thank the anonymous referees for their useful feedback on the original version of the manuscript. Additionally, we acknowledge the importance of equity and inclusion in this work and are committed to advancing such principles in our scientific communities.} KS is supported by an NSF Graduate Research Fellowship and by a Hertz Foundation Fellowship.  AL acknowledges support for this work by NASA through Hubble Fellowship grant \#HST-HF2-51363.001-A awarded by the Space Telescope Science Institute, which is operated by the Association of Universities for Research in Astronomy, Inc., for NASA, under contract NAS5-26555. This research used resources of the National Energy Research Scientific Computing Center, a DOE Office of Science User Facility supported by the Office of Science of the U.S. Department of Energy under Contract No. DE-AC02-05CH11231. 
\bibliography{refs}
\end{document}